# Observation of exciton-exciton interaction mediated valley depolarization in monolayer MoSe$_2$

Fahad Mahmood,[†] Zhanybek Alpichshev,[†] Yi-Hsien Lee,[‡] Jing Kong,[¶] and Nuh Gedik[*,†]

[†]Department of Physics, Massachusetts Institute of Technology, Cambridge, MA 02139, USA

[‡]Materials Science and Engineering, National Tsing-Hua University, Hsinchu 30013, Taiwan

[¶]Department of Electrical Engineering and Computer Science, Massachusetts Institute of Technology, Cambridge, MA 02139, USA

E-mail: gedik@mit.edu

Phone: +1 (617) 253-3420

### Abstract

The valley pseudospin in monolayer transition metal dichalcogenides (TMDs) has been proposed as a new way to manipulate information in various optoelectronic devices. This relies on a large valley polarization that remains stable over long timescales (hundreds of ns). However, time resolved measurements report valley lifetimes of only a few ps. This has been attributed to mechanisms such as phonon-mediated inter-valley scattering and a precession of the valley psedospin through electron-hole exchange. Here we use transient spin grating to directly measure the valley depolarization lifetime in monolayer MoSe$_2$. We find a fast valley decay rate that scales linearly with the excitation density at different temperatures. This establishes the presence of strong exciton-exciton Coulomb exchange interactions enhancing the valley depolarization. Our work highlights the microscopic processes inhibiting the efficient use of the exciton valley pseudospin in monolayer TMDs.

# Keywords

Transition metal dichalcogenides, valley polarization, 2D materials, valley dynamics, exciton interactions

Monolayer transition metal dichalcogenides (TMDs) have a 2D hexagonal lattice structure which breaks inversion symmetry. This results in a direct band-gap or valleys in the energy-momentum dispersion at the corners of the hexagonal Brillouin zone (i.e., the K and K' points). Strong spin-orbit coupling further splits the valence band and the carrier spin index becomes locked to the valley polarization.[1] Thus, excitations in a particular valley (K or K') can be generated via the optical selection rule. For example, circularly polarized light with a certain helicity will only generate excitations in either the K or K' valley but not both, as demonstrated by various optical experiments (e.g.,[1–3]). In this way, the valley polarization of excitonic quasi-particles (excitons, trions etc.) can be manipulated to process information in the emerging field of 'valleytronics'.[4]

The feasibility of these monolayer materials for this field is primarily determined by how quickly excitations generated in a particular valley depolarize. Initially, the valley depolarization time was predicted to be quite long ($\sim$ a few ns) by various photoluminescence (PL) measurements[2,3,5] based on the large degree of circular polarization in the emitted PL. However, a number of time-domain experiments such as transient Faraday[6] and Kerr rotation,[7–9] time-resolved photoluminescence (TRPL)[10–12] and transient reflection and transmission spectroscopy[6,13–16] have revealed valley polarization lifetimes orders of magnitude shorter ($\sim$ a few ps) than predicted. While there exist several theoretical proposals[16–19] to explain this, a consensus has yet to emerge on the exact mechanisms for the fast valley depolarization.

An understanding of these is crucial for engineering devices based on valleytronics.

The case for the monolayer TMD MoSe$_2$ is particularly interesting. Initially the amount of valley polarization in monolayer MoSe$_2$ was found to be < 5%, about 10 times smaller than in other monolayer TMDs.[20] However, other works[21,22] have recently found sizeable valley polarization (as high as 30%). This discrepancy is due to the excess energy in the system, i.e., the circular polarization of the emitted PL is maximized when the difference between the excitation energy and the A-exciton transition is minimized.[21] How this initially generated valley polarization in monolayer MoSe$_2$ decays as a function of time is so far an open question.

In this work, we use Transient Spin Grating (TSG) to directly measure the dynamics of valley depolarization in CVD-grown monolayer MoSe$_2$. Consistent with time-resolved measurements on other monolayer TMDs, we find a valley lifetime of a few ps at room temperature that increases to tens of ps upon cooling to 4 K. More importantly, we discover a fast valley depolarization rate that scales linearly with the exciton density indicating a two-particle process (e.g., exciton-exciton exchange interaction) involved in enhancing the valley depolarization. Such a mechanism has not been reported previously in literature. Lastly, the behavior of the fast depolarization rate with temperature further highlights the role of electron-hole exchange interaction in destabilizing the valley degree of freedom in monolayer TMDs.

TSG has been applied to a number of spin-split electronic systems to reveal spin relaxation times in quantum dots[23,24] and GaAs quantum wells,[25] spin helical modes[26] and spin-diffusion[27] in semi-conductor quantum wells. The TSG measurement of spin relaxation in randomly oriented colloidal suspensions of quantum dot[28] demonstrates its unique ability to measure valley relaxation times in CVD-grown monolayer TMDs which inherently have disoriented $\mu$m[29] size domains.

In the TSG technique (Fig. 1a), two linearly cross polarized pump laser beams interfere on the sample surface to generate a spatially modulated circularly polarized intensity across the excitation spot as typically understood by the Grating Decomposition Method.[30] Since opposite helicities of light excite particles with opposite valley indices, this generates a spatially modulated valley density of wavelength $\Lambda$ which is referred here as the 'valley grating'. The decay of this grating directly corresponds to a decrease in the local imbalance between the two opposite polarized valley excitations. This can happen either through valley depolarization or through spatial diffusion of the initially spatially separated K and K' excitations. Thus, the valley grating decay rate can be written as:[27]

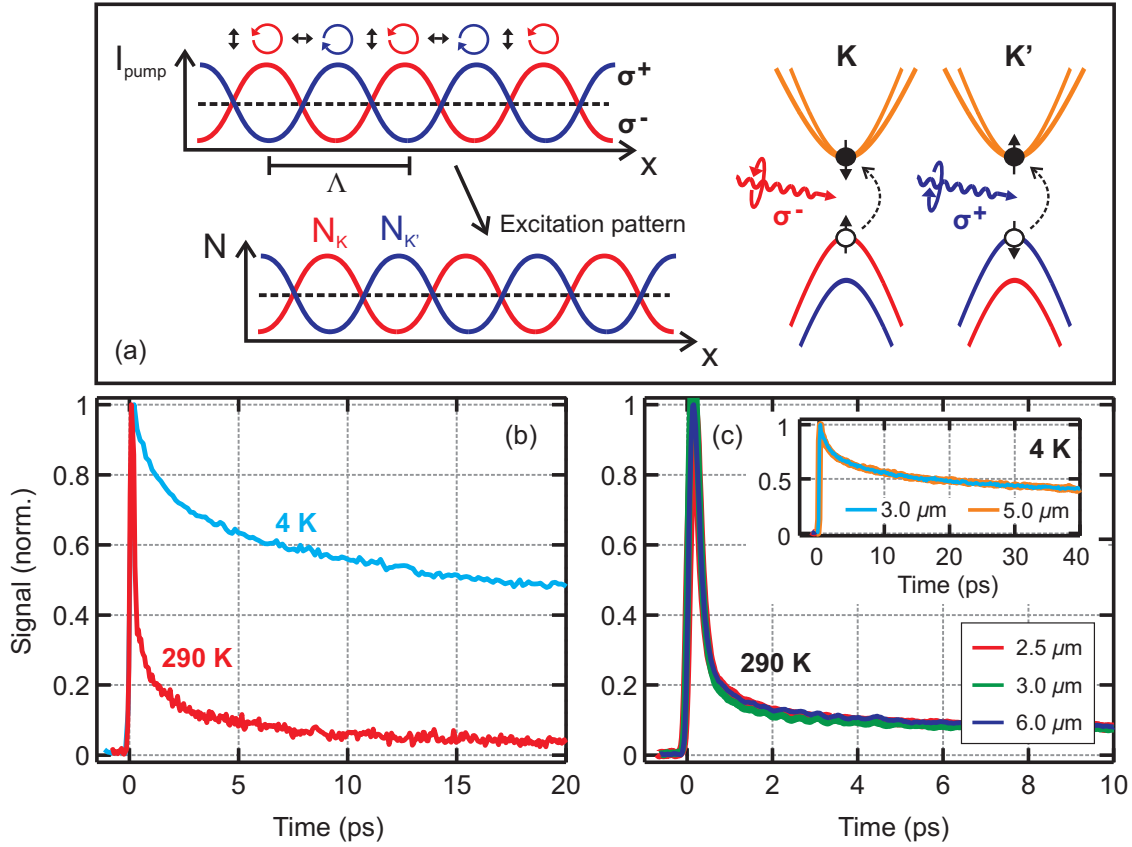

Figure 1: Transient spin grating (TSG) data on CVD-grown monolayer $MoSe_2$. **(a)** Two linearly cross-polarized pump beams interfere on the sample surface to generate a spatially varying circularly polarized intensity with wavelength $\Lambda$. Opposite helicities ($\sigma^+$ and $\sigma^-$) vary out-of-phase with each other on the sample (top-left). Due to the optical selection rule, opposite helicities excite carriers in opposite valleys (right). This leads to a spatially varying density of valley polarized excitations in the sample (bottom-left). **(b)** Intensity of the diffracted probe due to the valley grating as a function of the pump-probe delay time at both 290 K and 4 K. **(c)** TSG signal at 290 K for three different grating spacings $\Lambda$. Inset: signal at 4 K for two different $\Lambda$s.

$$\Gamma = D_v q^2 + \Gamma_v \tag{1}$$

where $D_v$ is the diffusion constant for the valley polarized excitations, $q$ is the modulus of the grating wave vector ($q = 2\pi/\Lambda$) and $\Gamma_v$ is the intrinsic valley depolarization rate. To measure the grating decay, a time delayed probe beam with linear polarization is incident on the valley grating and the intensity $I_v(t)$ of the resulting diffracted beam is detected with time (Fig. 1a). In our measurements both the pump and probe beams are set to an energy resonant with the A-exciton transition ($\sim 1.6\,\text{eV}$) in monolayer $MoSe_2$.[31] This ensures that the initial valley polarization is maximized while minimizing the effective excitation temperature. More details on the experimental technique can be found in the supporting information (SI).

Figure 1b shows the valley grating signal $I_v(t)$ as a function of the time delay between the pump and the probe at both 295 K and at 4 K. As can be seen, the grating decay is similar to the few ps valley lifetime reported by various time-resolved techniques[6,7,10–15] on monolayer TMDs. Moreover, the valley grating lifetime is much greater at 4 K when compared with room temperature. This could either be due to a decrease in the diffusion constant $D_v$ with decreasing temperature or a decrease in the valley depolarization rate. To separate out the effects of diffusion and valley relaxation, we measured $I_v(t)$ for various values of the grating wavelength $\Lambda$ (Fig. 1c). As can be seen, $I_v(t)$ is independent of $\Lambda$ at both 295 K and at 4 K indicating that the first term in Eq. 1 can be neglected. This is not surprising; since the mobility in CVD-grown films of TMDs is typically quite small due to the presence of traps,[32,33] the diffusion rate is expected to be negligible when compared to the relaxation rate. Therefore, by Eq. 1, the decay of the valley grating $\Gamma$ is a measure of the valley depolarization in $MoSe_2$ ($\Gamma_v$).

We now proceed to study this depolarization as function of the initial pump fluence i.e., on the initially excited particle density. While some studies on monolayer TMDs sug-gest an increase in the depolarization rate with increasing excitation density,[6] others report a negligible dependence.[7] Figure 2a shows $I_v(t)$ for various pump fluences ($\mathcal{F}$) ranging from $\sim 0.24\,\mu\text{J/cm}^2$ to $\sim 4\,\mu\text{J/cm}^2$. This corresponds to an exciton density varying from $2.64 \times 10^{10}\,\text{cm}^{-2}$ to $4.1 \times 10^{11}\,\text{cm}^{-2}$ (SI). With decreasing $\mathcal{F}$, the TSG signal slows down suggesting that the valley depolarization strongly depends on the excitation density. To determine whether this dependence is a feature of inter-valley scattering, we plot the optical pump-probe transient reflectivity $\Delta R/R$ of the MoSe$_2$ monolayer for the same pump fluences and energy as the TSG experiment (inset of Fig. 2a). This is known to encode the population dynamics of excitons[34–36] i.e., the relaxation of $N_k + N_{k'}$. As shown in the inset of Fig. 2a, $\Delta R/R$ turns out to be independent of the pump fluence indicating that the total population relaxation of valley excitations is independent of the initial density. This observation signifies that the fluence dependent decay channel in the TSG signal only effects the local difference in population of opposite valley excitations ($N_k - N_{k'}$); which is possible only through inter-valley scattering from the K valley to the K valley.

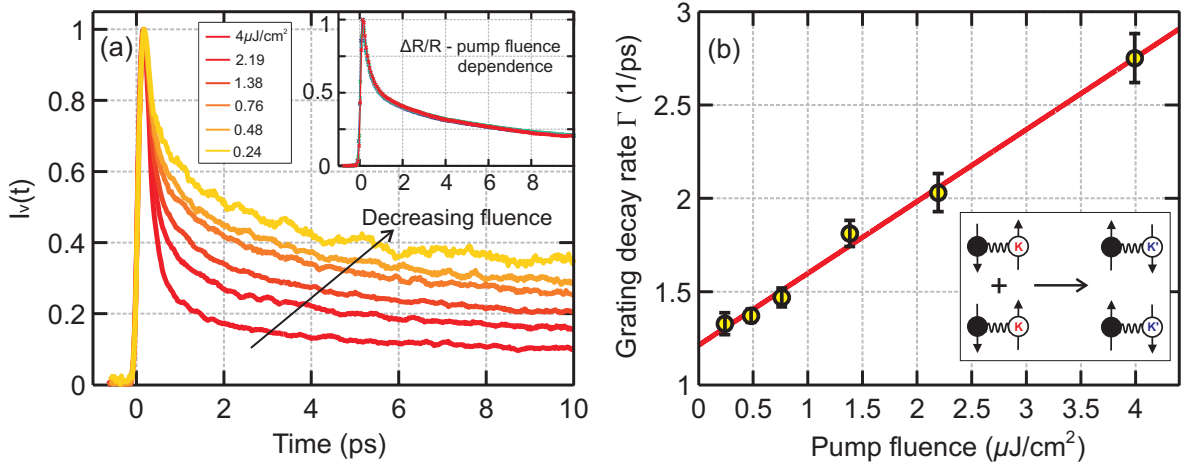

Figure 2: Effect of excitation density on the valley depolarization. **(a)** TSG signal at different pump fluences. Inset: Fractional change in reflectivity ($\Delta R/R$) of mono-layer MoSe$_2$ for various pump fluences. The data are obtained simultaneously as the TSG signal using a heterodyne detection scheme (SI). **(b)** Grating decay rate (fast component) as a function of the pump fluence. The fast rate at each fluence is obtained from a fit to the TSG signal in (a) (SI). Error bars represent the 95% confidence interval (2 s.d.) in the fitting parameters. Red line is a linear best fit to the data points. Inset: Schematic illustrating the bimolecular process involved in the valley depolarization.

In order to further study the fluence dependent decay, we fit the TSG data at each fluence to a sum of two exponentials to extract a 'fast' and 'slow' decay rate (SI). Figure 2b shows that the fast rate ($\Gamma_v$) scales linearly with pump fluence $\mathcal{F}$ whereas the slow rate is independent of the fluence (SI). For the purpose of this study, we focus on understanding the fast rate since it dominates the initial valley decay and the observed linear dependence on fluence has not been reported before for valley depolarization in any monolayer TMD. One trivial reason for the valley depolarization getting faster with increasing fluence is an increase in the effective temperature of the excitations. However, since we excite the system at the A-exciton resonance, the change in the effective temperature is determined by the pump bandwidth rather than the fluence. In addition, $\Gamma_v$ remains linear with fluence for all external temperatures from 3.5 K to 290 K (Fig.4 a), further ruling out an increase in the effective temperature as explaining the observed linear fluence dependence.

Rather, this type of linear dependence is typically seen in quasi-particle recombination in high-$T_c$ superconductors[37] as well as in exciton-exciton annihilation in various semiconductors including monolayer $MoS_2$.[35] In such bimolecular processes, two exited particles interact with each other to induce a decay in their overall population. Thus, for valley polarized excitons in $MoSe_2$, $\Gamma_v$ being proportional to $\mathcal{F}$ suggests that two excitons with the valley index (K) interact with each other to produces two excitons with the opposite valley index (K') (inset Fig. 2b).

We interpret this result in the context of the Maialle-Silve-Sham (MSS) mechanism which has been proposed by T. Yu et. al.[19] to explain the efficient valley depolarization in monolayer TMDs. In this mechanism excitons with opposite valley polarizations are assigned opposite valley pseudospins. The electron-hole exchange interaction within an exciton provides a a momentum ($\vec{k}$) dependent magnetic field $\Omega(\vec{k})$ around which the valley pseudospin can precess. Excitons with different center-of-mass precess with different frequencies. Thus, any random momentum scattering of the excitons will influence the overall valley depolarization rate. Based on the typically large impurity concentration in CVD grown TMD samples (SI), we conclude that our system is in the strong scattering regime: the momentum scattering rate $\tau_k^{-1}$ is much greater than the precession frequency i.e., $\tau_k^{-1} \gg \left\langle |\Omega(\vec{k})| \right\rangle$. In this regime, the momentum of an exciton changes continuously due to various scatterers (e.g., impurities, phonons, excitons) before the valley pseudospin can complete a full precession. Thus, similar to electron spin decay via the D'yakonov-Perel (DP) mechanism,[38,39] the valley depolarization rate scales inversely with the momentum scattering rate:

$$\tau_v^{-1} \propto \langle \Omega^2(\vec{k}) \rangle \tau_k \qquad (2)$$

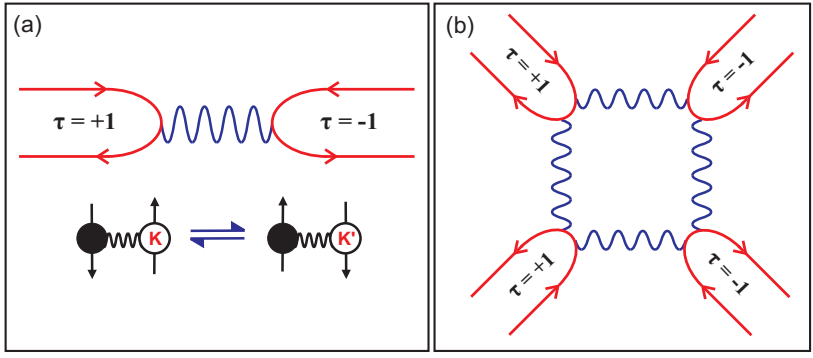

Figure 3: Electron-hole exchange processes due to Coulomb interactions **(a)** within an exciton and **(b)** between two excitons. $\tau$ denotes the valley index.

While this mechanism correctly predicts a valley polarization time of a few ps instead of ns,[19] it does not fully explain our observed fluence-dependent valley depolarization. In fact, according to Eq. (2), with an increase in exciton density i.e., an increase in momentum scattering due to exciton-exciton collisions, one would naively expect a decrease in the valley depolarization rate. However, we observe the exact opposite (Fig. 2a). Moreover, given that the estimated impurity concentration ($\sim 10^{14}\,\text{cm}^{-2}$) (SI) is much greater than the exciton density ($\sim 10^{11}\,\text{cm}^{-2}$), the momentum of an exciton is more likely to change due scattering with an impurity rather than through exciton-exciton scattering. Thus, Eq. 2 cannot account for our observed fluence dependence.

To explain this discrepancy, we note that previous literature on the MSS mechanism has only considered e-h exchange interaction within an exciton. This leads to an annihilation of a K-exciton to yield a K'-exciton as illustrated in Fig. 3a. However, we can also consider the interaction between two K-polarized excitons which will then lead to the annihilation of two K-excitons to generate two K'-excitons as shown in Fig. 3b. This exciton-exciton interaction mediated valley depolarization scales linearly with the exciton density and thus is a probable explanation for our observed fluence dependence.

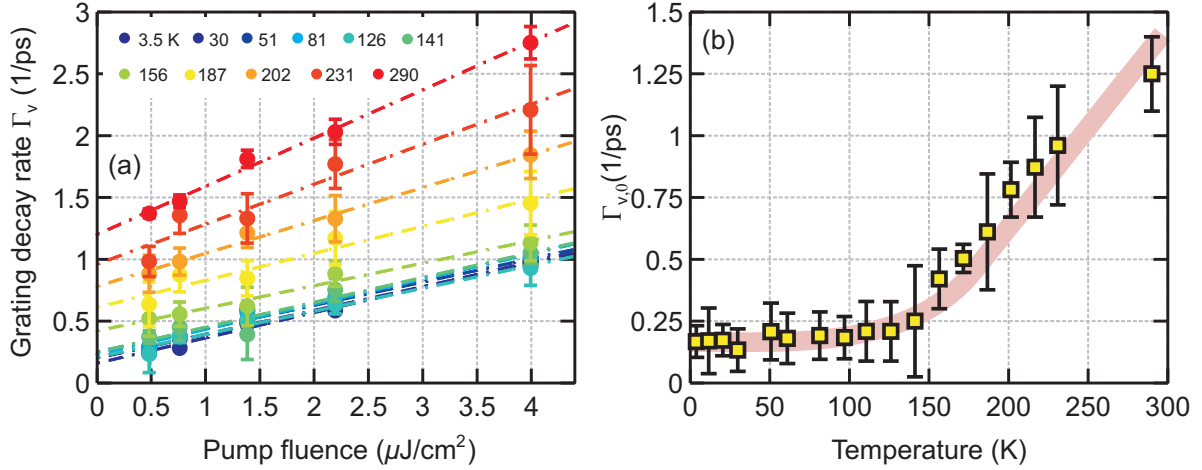

Figure 4: Effect of temperature on the valley depolarization. **(a)** Grating decay rate (fast component) as a function of pump fluence for various temperatures. Straight colored lines are linear best fits to the data at each temperature. **(b)** $\Gamma_{v,0}$, obtained from the y-intercept of the linear best fits in (a) as a function of temperature. Error bars represent the 95% confidence interval (2 s.d.) in the fitting parameters. Pink line is a guide to the eye.

If the fast valley depolarization rate $\Gamma_v$ was just determined by exciton-exciton exchange interaction, then $\Gamma_v$ should go to zero as the pump fluence $\mathcal{F} \to 0$. However, $\Gamma_v$ vs $\mathcal{F}$ in Fig. 2b has a positive non-zero intercept. Thus, $\Gamma_v$ can be separated as: $\Gamma_v = \Gamma_{v,0} + \beta n$, where $\beta$ is a phenomenological constant, $n$ is the valley polarized exciton concentration (set by the fluence) and the non-zero intercept $\Gamma_{v,0}$ is due a decay channel different from exciton-exciton exchange interaction. Based on previous studies[7] in monolayer TMDs, we posit that e-h exchange interaction within an exciton, as described above for the MSS mechanism, is a likely candidate to explain $\Gamma_{v,0}$.

To understand this further, we took TSG data on monolayer $MoSe_2$ at various different temperatures. We extract the fast depolarization rate and plot it as a function of pump fluence in Fig. 4a. The rate scales linearly with fluence for all temperatures with very little change in the slope ($\beta$). $\Gamma_{v,0}$ is then extracted from the y-intercept for each trace and the resulting values are plotted as a function of temperature on Fig. 4b. $\Gamma_{v,0}$ is constant at low temperature and increases roughly linearly with temperature for $T > 130\,\text{K}$. This behavior can indeed be explained as part of the MSS mechanism. If the homogenous or collisional broadening of excitons is greater than their average thermal energy ($\hbar/\tau_k > k_B T$), then the valley depolarization rate should be independent of temperature.[7,18] Here $\tau_k$ is the average momentum scattering time. Thus, the low temperature behavior of the valley depolarization is determined by the collisional broadening of the A-exciton. As the temperature increases, the thermal energy becomes comparable to collisional broadening ($\hbar/\tau_k \sim k_B T$) and the resulting thermal variation in th effective magnetic field scales linearly with the temperature i.e., $\langle \Omega^2(\vec{k}) \rangle_T \propto T$. This causes thermal fluctuations in the precession frequency of the valley pseudospin and thus the valley depolarization rate $\Gamma_{v,0} \propto T$ at high temperatures which is consistent with the data.

We have also considered the possibility of $\Gamma_{v,0}$ originating due to phonon-mediated valley depolarization. This mechanism has been invoked to explain the ns long decay of the pseudospin of resident holes in monolayer TMDs (e.g., Hsu et. al.[40]). In that case, the decay rate is much smaller than what we observe for $\Gamma_{v,0}$ but still follows a similar temperature dependence. Indeed, using a similar model for phonon occupation, we can describe $\Gamma_{v,0}$ with temperature using a phonon energy of $E_p = 37 \pm 9\,\text{meV}$ (SI), which interestingly is quite close to the measured energies of an optical $E_{2g}^1$ (35.3 meV) phonon in monolayer MoSe$_2$.[41,42] It is also roughly twice the energy of the longitudinal acoustic phonon at the zone edge (19 meV).[41] This suggests that zone edge phonons might be involved in determining $\Gamma_{v,0}$. However, the fast rates we measure are an order of magnitude greater than what is typically expected for phonon-mediated valley depolarization from both theory[19,43] and experiment.[40] It may be that 'slow' component in the observed bi-exponential valley decay (SI) is explained by a phonon-mediated process and that can be addressed in future TSG experiments that resolve the signal upto ns time-scales.

Based on the above discussion we conclude that e-h Coulomb exchange interaction both within and between excitons dominates the observed fast valley depolarization, with the latter increasing the valley decay with increasing exciton density. Future work will consist of TSG experiments on gated devices of $MoSe_2$ to study the effect of charged excitations (e.g., trions) on the valley polarization decay. Similarly, these experiments can also be carried out in a magnetic field similar to time-resolved Kerr rotation measurements on $WSe_2$.[44] The resulting Zeeman splitting of the conduction band is expected to modify the electron-hole exchange interaction and thus modify the valley depolarization time. We note that our work is based on CVD grown samples which inherently have larger impurity concentrations than exfoliated samples. It is possible that clean exfoliated samples of monolayer $MoSe_2$ with high mobilities display different behavior due to the system being in the weak scattering regime of the MSS mechanism explained above.

# Acknowledgement

The authors thank Furkan Cagri, Alex Frenzel, Liang Fu, Joe Orenstein, Edbert Sie, Darius Torchinsky and Di Xiao for useful discussions. This work is supported by the U.S. Department of Energy, BES DMSE (experimental setup and data acquisition), and from the Gordon and Betty Moore Foundation's EPiQS Initiative grant GBMF4540 (manuscript writing). YHL acknowledges support from AOARD grant (co-funded with ONRG) FA2386-16-1-4009, Ministry of Science and Technology (MOST-105-2112-M-007-032-MY3 and MOST-106-2119-M-007 -023 -MY3).
The authors declare no competing financial interest.

# Supporting Information Available

Supporting Information.pdf: includes further experimental and analysis details with 3 additional figures.